# Design And Develop Network Storage Virtualization By Using GNS3

Abdul Ahad Abro, Ufaque Shaikh

*Abstract* — Virtualization is an emerging and optimistic prospect in the IT industry. Its impact has a footprint widely in digital infrastructure. Many innovativeness sectors utilized the concept of virtualization to reduce the cost of frameworks. In this paper, we have designed and developed storage virtualization for physical functional solutions. It is an auspicious type of virtualization that is accessible, secure, scalable, and manageable. In the paper, we have proposed the pool storage method used the RAID-Z file system with the ZFS model which provides the duplication of site approach, compression blueprint, adequate backup methods, expansion in error-correcting techniques, and tested procedure on the real-time network location.  Therefore, this study provides useful guidelines to design and develop optimized storage virtualization.

## I. INTRODUCTION

Virtualization and network storage management are essential issues in the development of data centers. The assignment of a virtual machine to physical machine impacts the model and the idle time of physical machines. In virtualization, the default value is extensive including server consolidation, network maintenance, stable computing, kernel debugging, and device migration. Most have a common operating environment for the end-user, but also very different levels of architecture and service abstraction. With the increase in popularity and benefits of virtualization, most of the organizations are shifting their business to virtualization.

Moreover, the Issue of information security of sensitive data stores, in particular data integrity. Organizations are still vulnerable to silent manipulation in data [1].  Server consolidation, fault tolerance are the key factors to improve the performance in the cloud virtualization [2]. The aggregation of servers minimizes resource consumption by raising the number of servers operating in the cloud. To reduce power consumption in the cloud, virtual machine instances of underused servers are migrated to the ideally suited servers [2].  Load management distributes the fee across the servers. The virtual machine instances of highly charged servers are moved to poorly charged servers [2].

## II. RELATED WORK

Several methods have been discussed for the network storage virtualization, and thus this section discusses some of the techniques using GNS3.

In the related work section, NetStorage which aims to synchronously replicate processing and trace networks on target devices. NetStorage's aim is to restore the same network and processing workloads over different times as initial applications of network storage systems [3]. ZFS combines a conventional file system and conceptual capacity controller functionality and uses a shared storage model to promote a dynamically file system that supports versatile and large disk arrays. The connecting block devices to the storage list, ZFS calls Zpools [1]. Generic RAID reassembly using system entropy provides a novel method for the automated generic notification of all applicable RAID specifications using block-level entropy and standard heuristics [4]. Our experiments compare the performance concerning Zero downtime, storage space buffer, CPU, network access, use of switches, external memory, device workload, and usage of processors. The data center is a series of many services such as network devices, cryptographic algorithms, and computational devices. Cloud computing uses the data warehouse network to provide connectivity to computational power for the programs that the system needs [5]. The key performers in companies such as Google, Yahoo, and Microsoft have also introduced this concept in their private clouds, supplying millions of users with a variety of services [6][12]. In optimizing live virtual machine relocation in the cloud-centered on redundancy, it may save bandwidth utilization and the overall migration period by avoiding the migration of similar storage sites [7].

## III. VIRTUALIZATION

### A. Operating System Virtualization

A virtual system or hybrid network has been a sizeable thing of correspondence innovation engineering. A virtual computer is a normal working framework along with Windows Vista or Red Hat Enterprise Linux which works as the complete usage of the equal bodily gadgets. Virtual Machine Manager (VMM) directed the digital machines in an unexpected way, and each case of the working framework (OS) executes and handles the capability proficiently and viably. Corporations like Microsoft, Intel, VMware, and AMD contribute to the splitting up of the operating system through relationships. The key mechanism and advocate of data centers for taking the

Abdul Ahad Abro, Department of Computer Engineering, Ege University, Turkey. Email: 91150000622@ogrenci.ege.edu.tr1

Ufaque Shaikh, Department of Computer Science, Ilma University, Pakistan. Email: ufaqueshaikh@gmail.com

rewards of the traditional virtual machines industry. In this way, companies use the range of physical devices to increase their data centers and decrease the number of requests. This cycle proceeded in the role of equipment, rack spaces, controls, and cable management.

B. Application Virtualization

In this phase, the concept of a similar standardized application, virtual machines, and servers are entirely different. The applications are clients by default. The Microsoft SoftGrid Virtualization Application is a good example. The Microsoft Word distributed on-site and personal information held in the state and SoftGrid bilateral administration and deployment. It includes the machine and memory required for running the program but nothing is mounted on the computer globally. These applications depend on virtual run frameworks which execute in local and remote application management. In hybrid forecast-based virtual machine relocation, virtual machines are transferred from servers that struggle to satisfy the condition of load balancing to chosen endpoint servers using a load-conscious migration algorithm.[8].

C. Network Virtualization
A virtual private network (VPN) is as a matter of direction putting in frameworks. The degree to which digital IP replication is transmitted and dealt with is VLAN IP: an Ethernet port that can oblige a number of IP addresses and a few virtual systems. The virtual IP has one physical port contact independently and would not understand every other's essence, however modifications the connection independently and handles them. Furthermore, the proportion of IP guidance tables and server ports is 01:01, however there are some virtual interfaces organizing, (for example, port or VLAN just as ETH0). Linux is regular with the virtual system connector. The virtual mannequin sending table to modify and holds the extent of a few guidance tables with a range of passages for each real bodily gadget. This interface gives simple directing gadgets without wrecking system benefits and steerage tables internal a similar structure.

D. Hardware Virtualization
Virtual machines (VM) are very close to the virtualization mannequin of the operating system. To distribute the VM bodily infrastructure into person components and to discover a single element as a separate section. In this case, the time and configuration cycle does now not function on processing, the OS simply re-CPUs the scheduling functions, which are responsible for gadget time allocation.

E. Storage Virtualization
The digital square level virtualization is the fashionable verified innovation. It is the accompanying widespread condensed restrict the storage area network (SAN) and network attached storage (NAS) innovation. System stockpiling restrict seems, through all accounts, to be a specific lightweight bodily gadget. In this process, SAN devices often form the storage of virtualization. The ISCSI protocol is commonly achieved, and another standard is set for OS application of a virtual device, including LAN adapter (software or hardware) deployment and the physical existence of the command device. ISCSI mechanism in the network adapter is to pass the SAN extensive call blocking program kit and transfer it to the virtual disk. In resource-conscious virtual machine relocation, the servers are classified into eight clusters based on the CPU, memory usage, and work arrival rate. The priority attribute is used to classify the crowded servers. Using a resource-aware virtual machine migration strategy, the virtual machines from the server with high priority values are relocated to the target server.[9]. Virtual machine transition acts as a crucial method for cloud virtualization to tackle workload management, data reduction, failure tolerant, and hardware repair scenarios[10]. In virtual machines, finding the optimum compromise around host utilization and energy consumption, the suggested architecture assures that host computers run even the most control-efficient rates of utilization, i.e. the rates with the maximum PPR, thus the energy consumption considerably with the dismissable output.[11][13]. Virtualization services, application, data source, OS, hardware, and physical storage are illustrated in fig 1.

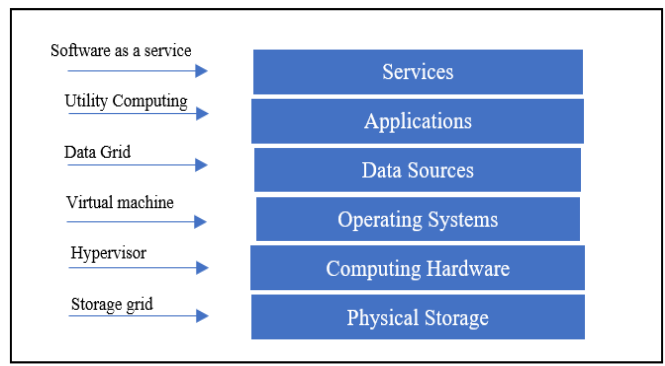

Fig 1: Distribution of Virtualization Layers

IV. METHODOLOGY
In this study, each time new records are deposited in the file system and the analysis of the overview of the current data pool. Meanwhile, these backups repeat on the local NAS and then synchronize the local NAS with the cloud NAS utilizing altered backup methods such as discrete or cumulative, which ZFS embraces. The NAS uses FreeNAS, the network's FreeBSD Linux platform. Throughout this case, we suggested building a simulated monitoring network for the GNS3, as outlined in fig 2. Virtual machine efficiency measured via the virtual package, a free resource that oracle offers for virtualization of OS. The manufacturer launched Sun

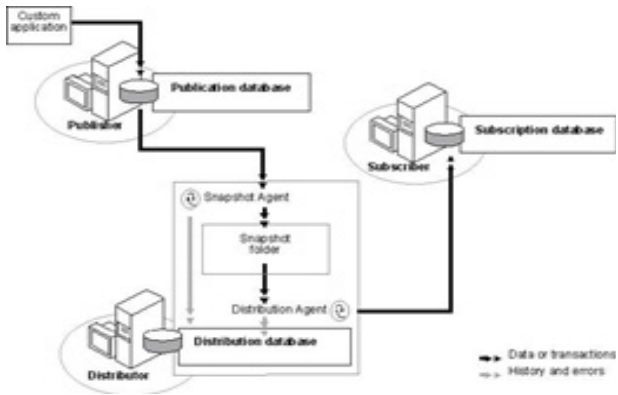

Fig 2: Virtual Box Scenario

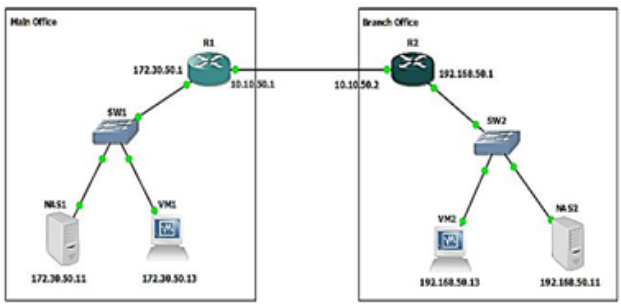

Fig 3: Virtual Machines with Virtual Network Environment

Microsystem's Zeta Byte File System. To sum up, the subscriber shows the BSD system, which is the version of UNIX.

## V. TOPOLOGY

This set of VirtualBox is integrated with GNS3, which utilized the virtual machines in the existence of a virtual network environment. In this process, we have proposed an approach utilized the four virtual machines (VM) for OS, two storages, routers, and switches, which are used in an experimental environment. The connectivity of this network is highly scalable, which has the capacity to enhance additional sites accompanied by further configuration. During this process, some concerned parameters have been used for smooth transmission and error-free, which is network access, up-time, storage capacity collection, swap use, physical memory, device load CPU and storage use.

## VI RESULT AND DISCUSSION

The performance of CPU usage demonstrates that the amount of inactivity, code execution, and implementation of client code in various countries, which is illustrated in fig 4. In fig 5, system load, which shows the average fifteen minutes utilizing the approach. The disk shows the mean period of hard disk drive I/O in fig 6. Physical memory in fig 7, indicates

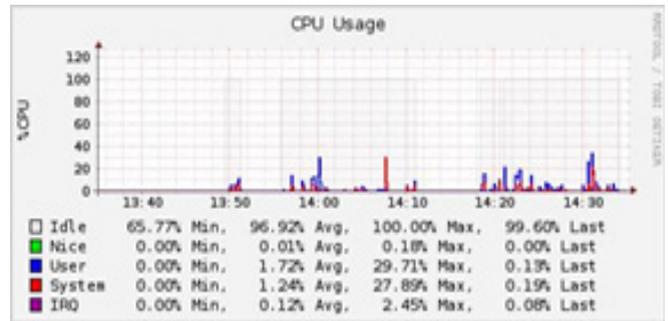

Fig 4: CPU Usage

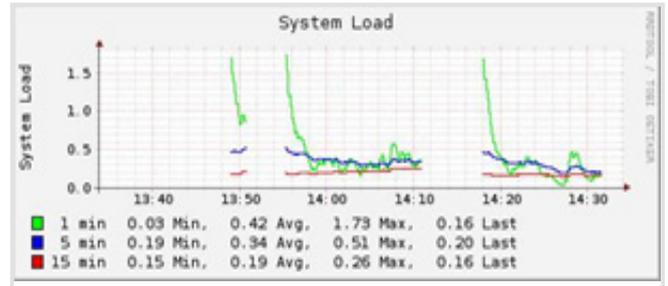

Fig 5: System Load

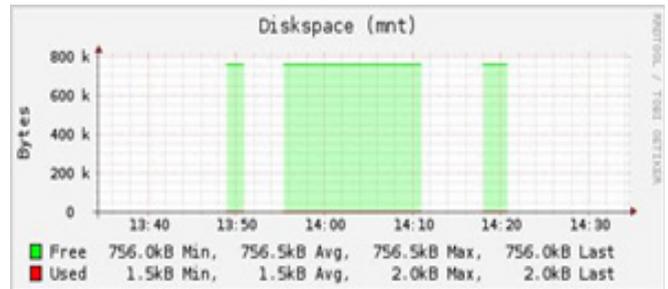

Fig 6: Diskspace

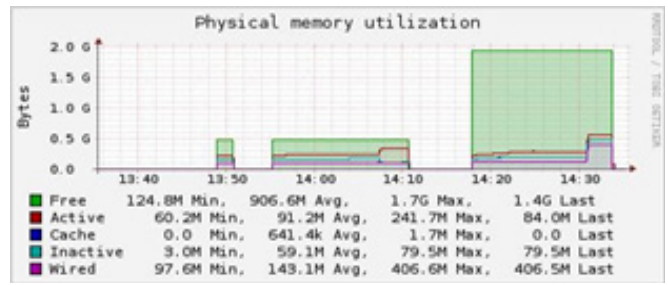

Fig 7: Physical utilization

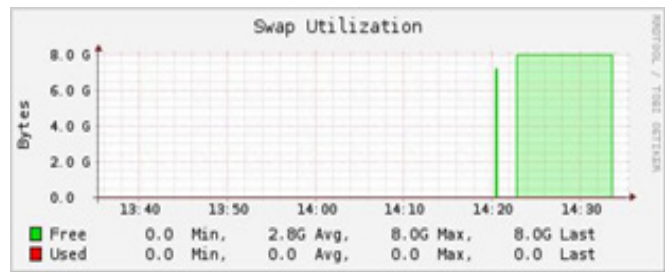

Fig 8: Swap Utilization

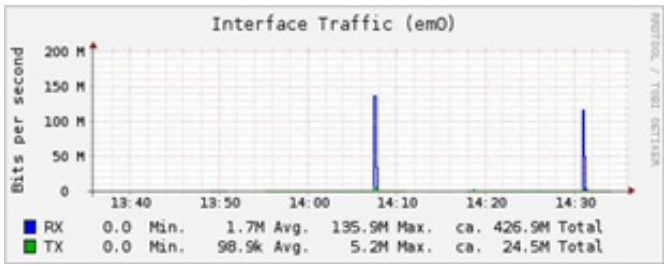

Fig 9: Interface Traffic

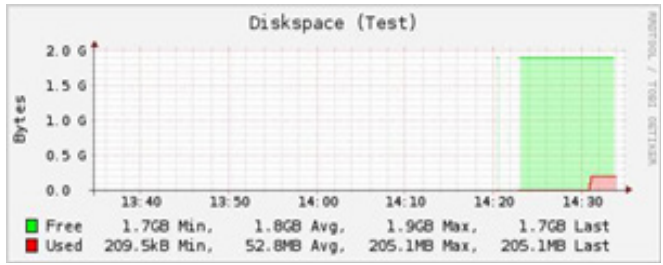

Fig 10: Diskspace

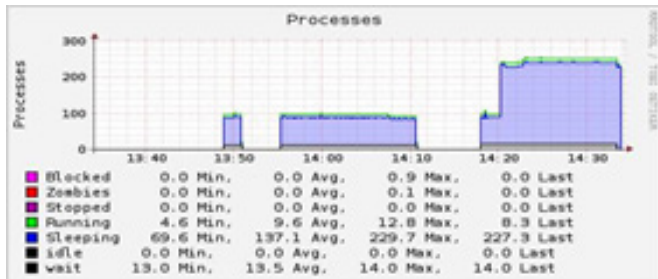

Fig 11: Processes

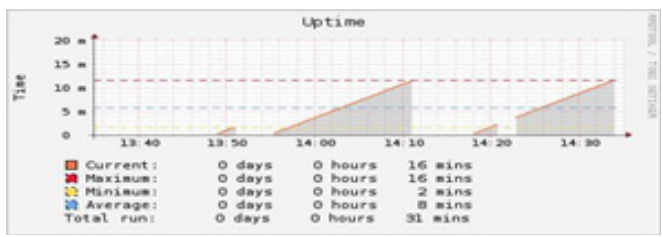

Fig 12: Uptime

the usage of NAS. In fig 8, presented the number of swap space accessible to exhibit.

Moreover, in fig 9, represented the received the sum of data transmitted per second configuration bits per-view framework. Furthermore, the fig 10, disk space is demonstrated for each capacity and available space, and the data set. However, in fig 11, the process phase pointed out the numeral of methods, collected by state-owned. Lastly, in fig 12, the uptime standard processing time to concentrated accessibility of the approach.

## VII. CONCLUSION

Z File System (ZFS) is regarded the consolidated, dynamic, functional, social semantics, begin to end archive security, and unmatched adaptability. ZFS is certifiably now not a dynamic overhaul of current advances. It's the shape of the utility framework. ZFS gives in the a variety of varieties of the framework sturdiness, which is named RAID-Z. This resembles RAID-5, yet with dynamic transfer speed to RAID-5 to shut the void recorded as a tough copy by using evacuating code spillage usefulness and focusing on consistency changes. The ZFS consistently previews and rehashes a steady. The read-just preview rendition of the report framework is a 2d in time when enjoying out a depiction writable chronicle.

It has restores and backup capture functionality. The growing shot will create a complete archive and produce a progressive backup of the snapshot. It is practical, which can be used to remotely reproduce gradual change every 10 seconds. Moreover, some exceptional feature is: It supports the large database (16 Exabytes or 18 million Terabytes. An auto-recovery file structures (ZFS works 256-bit checksums end-to-end to authenticate data).

In the future, we will check it on the actual circumstance by building memory pool over various sites and modifying dedicated hard disk utilizing specific circumstances like synchronization mode adjustment, there will also be a lot of research on error fixing strategies and encoding methods that improve gradual sample replication of operational skills.